\documentclass[traditabstract]{aa}

\usepackage{graphicx}
\graphicspath{{../../images/}}
\usepackage{txfonts}

\usepackage{epsfig}
\usepackage{natbib}
\newcommand{\beq}{\begin{equation}}
\newcommand{\eeq}{\end{equation}}

\makeatletter 
\def\eqalign#1{\null\,\vcenter{\openup\jot\m@th  \ialign{\strut\hfil$\displaystyle{##}$&$\displaystyle{{}##}$\hfil      \crcr#1\crcr}}\,} 
\makeatother

\begin{document}

\title{Modelling ion populations in astrophysical plasmas: \\ carbon in the solar transition region.}
\author{R. P. Dufresne
\and
G. Del Zanna 
}
\institute{DAMTP, Univerity of Cambridge, Wilberforce Road, Cambridge CB3 0WA, UK}

 \date{Received  ; accepted }

 \abstract{The aim of this work is to improve the modelling of ion populations in higher density, lower temperature astrophysical plasmas, of the type commonly found in lower solar and stellar atmospheres. Ion population models for these regions frequently employ the coronal approximation, which assumes conditions more suitable to the upper solar atmosphere, where high temperatures and lower densities prevail. These assumptions include all ion charge-states being in the ground state, and steady-state equilibrium where there is sufficient time for ionisation and recombination to take place. Using the coronal approximation for modelling the solar transition region gives theoretical lines intensities for the Li-like and Na-like isoelectronic sequences which can be factors of 2-5 times lower than observed. The works of Burgess \& Summers (1969) and Nussbaumer \& Storey (1975) showed the important part ions in excited levels play when included in the modelling. As density increases metastable levels become populated and ionisation rates increase, whereas dielectronic recombination through the highly-excited levels becomes suppressed. Photo-ionisation was also shown by Nussbaumer \& Storey to have an effect on the charge-state distribution of carbon in these regions. Their models, however, used approximations for the atomic rates to determine the ion balance. Presented here is the first stage in updating these earlier models of carbon by using rates from up-to-date atomic calculations and more recent photo-ionising radiances for the quiet Sun. Where such atomic rates are not readily available, in the case of electron-impact direct ionisation and excitation--auto-ionisation, new calculations have been made using the Flexible Atomic Code and Autostructure, and compared to theoretical and experimental studies. The effects each atomic process has on the ion populations as density changes is demonstrated, and final results from the modelling are compared to the earlier works. Lastly, the new results for ion populations are used to predict line intensities for the solar transition region in the quiet Sun, and these are compared with predictions from coronal-approximation modelling and with observations. Significant improvements in the predicted line intensities are seen in comparison to those obtained from zero-density modelling of carbon.}
 	

\maketitle

\section{Introduction }
\label{sec:intro}

This paper is the first in a series of studies in which the aim is to improve the modelling of the transition region (TR) lines, that is, those lines emitted by the thin layer between the chromosphere and corona \citep[][]{mariska1992}. One long-standing issue regards the strongest lines in the UV and EUV from the Li- and Na-like ions, which are significantly enhanced, by factors ranging from two to five, compared to their predicted values, as obtained using lines from other ions. Notable examples are the \ion{C}{iv} and \ion{Si}{iv} lines, the latter being routinely observed by the IRIS satellite \citep{depontieu_etal:2014}. This characteristic behaviour of the Li- and Na-like ions was first noted by \cite{burton_etal:71}, although even earlier observations present the same problem (see the Solar Living Review by \citealt{delzanna_mason:2018}). \cite{delzanna_etal:02_aumic} showed for the first time that the same problem occurs in stellar transition regions. This is an important issue for astrophysics, not just for solar physics. 

Non-equilibrium effects have long been thought to be the most likely cause. For example, departures from ionisation equilibrium can enhance significantly some of the ions that have long ionisation and recombination times, as shown in \cite{bradshaw_etal:04} for example, or in the case of IRIS lines by \citet{Olluri13b} and \citet{Martinez16} for instance. Non-Maxwellian electron distributions tend to shift the formation temperature of the TR lines towards lower values, (see the review of non-equilibrium processes by \citealt{dudik2017},) leading to an enhancement in predicted intensities, as shown for the \ion{Si}{iv} case by \cite{dudik_etal:2014_o_4}. 

However, there are several other physical processes which could affect the modelling of these ions and which are usually not taken into account in the literature, \citep[see the review by][]{delzanna_mason:2018}. The main ones discussed in the present paper are density effects on the charge-state distribution through electron-impact ionisation and dielectronic recombination. The r\^{o}le of photo-ionisation on the charge-state distribution is also explored. 

\cite{burgess1969} were the first to point out density effects on the ion balance, principally through the suppression of dielectronic recombination. This was followed by \citet{nussbaumer1975}, who incorporated the effects of both density and photo-ionisation in a collisional-radiative (CR) model for carbon. The importance of the metastable levels with density in the ionisation balance was expanded by \citet{mcwhirter1984}, followed by a CR modelling suite made available to the Atomic Data and Analysis Structure (ADAS) consortium, later described by \citet{summers2006} for example. Some density effects were included approximately in modelling by \cite{vernazza1979}, and in modelling solar irradiance measurements by \cite{judge1995}, although the latter concluded that density effects alone could not account for the large discrepancies in the intensities of Li-like and Na-like ions. More accurate modelling of the density effects have used ADAS rates, \citep[see e.g.][]{doyle_etal:2005}. However, in most atomic databases \citep[e.g. Chianti\footnote{www.chiantidatabase.org},][]{dere1997, delzanna_chianti_v8} and literature, density effects and photo-ionisation are usually neglected. Charge-state distributions are used for a wide range of diagnostics, and thus it is important that density-dependent ion populations are used when interpreting lines from the TR, such as for the measurements of densities \citep[see e.g.][]{polito_etal:2016b} or chemical abundances (as discussed e.g. by \citealt{young2018}).

The aim here is to build a database of new atomic data and codes to take into account these and other effects which are not normally included in modelling. The ultimate goal is to study all the main effects in a self-consistent way, through several steps. To begin this process, carbon is discussed in this paper, as all the main physical processes were included by \cite{nussbaumer1975}. In most cases the atomic rates available at the time were very approximate; more accurate rates for most of the relevant processes have been published since their work. The principal exception is collisional ionisation by electron impact, especially from the excited states, which is the main focus of this paper, along with building the first stages of a CR model. In addition, the effect of photo-ionisation is explored, since this was shown by \citeauthor{nussbaumer1975} to be an important process for carbon in the lower TR, although it is noted that other effects, such as opacity, are at play here. There is currently a renewed interest in modelling the low charge-states of carbon, such as singly-ionised \ion{C}{ii} \citep[see e.g.][]{rathore_carlsson:2015}, as spectral lines are routinely observed by IRIS and used to study this part of the solar atmosphere.

The next section presents an overview of the relevant atomic processes applicable to this region, and the processes for which new calculations have been made. In addition, it includes a discussion about how ion populations are calculated. The results of both the atomic calculations and ion population modelling are presented in Section \ref{sec:results}, as well as a comparison with previous work. The impact that the new results have on the predicted line intensities and how they compare to observations of the quiet Sun are demonstrated in Section \ref{sec:obs}, and brief conclusions and the outlook for further work are given in the final section.

\section{Methods}
\label{sec:method}

As mentioned above, for solar TR densities of $N_{\rm e} \approx 10^{10}~{\rm cm}^{-3}$ there are two atomic processes which are particularly important in shifting the formation temperature of an ion when compared to zero-density conditions. The first one is electron-impact ionisation from excited levels of the lower ionisation stages, and the second is suppression of the dielectronic recombination rates. 

In plasmas at these densities, ions are collisionally excited and a significant proportion occupy metastable levels just above the ground. The levels last sufficiently long that ionisation can occur from these levels, \citep[see for instance][]{mariska1992}. Ionisation rates from those levels are faster than from the ground, and the ensuing result will be greater populations in the next higher charge-state compared to when metastable levels are not present in the modelling.

Dielectronic recombination (DR) was shown by \cite{burgess:64, burgess:65} and \cite{seaton:64} to be the main recombination process for the solar TR and corona. Recombination at these temperatures for this process takes place into high-lying levels. Recombined ions in these states will be rapidly re-ionised by electron collisions, again causing larger populations to exist in a higher charge-state than would be present when high-lying levels are not included. Following the discovery of the importance of DR in the solar atmosphere and the first ionisation-equilibrium calculations which included it \citep{burgess_seaton:1964}, many calculations have been published over the years \cite[e.g.][]{arnaud_rothenflug:85,mazzotta_etal:98}. However, the vast majority are based on calculations at zero electron density. Indeed, even at the lower density of the quiet solar corona ($N_{\rm e}=10^{8}$ cm$^{-3}$) some density-dependent effects can be present.

Many codes and CR models have been developed for atomic fission or fusion, but are not publicly available. The original CR model of \citeauthor{burgess1969} was developed further by \cite{summers:1972,summers1974}, and then implemented within ADAS. In most cases, the ion models were simplified by bundling levels into either LS terms, configurations or bundles of n-resolved levels. The results of the CR modelling have recently been made available as {\it effective} ionisation and recombination rates via OPEN-ADAS\footnote{open.adas.ac.uk}. Several inconsistencies in the OPEN-ADAS rates have recently been noted \citep{delzanna2019}. The reasons for the inconsistencies are difficult to ascertain, as details of the basic atomic rates used for the modelling are not available. 

The objective, then, is to build a CR model based on up-to-date, level-resolved rates, where available. The main atomic processes necessary to model the plasma conditions of the solar TR are: ionisation by electron impact, photo-ionisation, radiative and dielectronic recombination, collisional excitation by electron and proton impact, and radiative decay. Because a full CR model including all the high-lying levels is quite complex on its own, the first stage has primarily been to assess the effect of collisional ionisation from the metastable levels and the suppression of DR on the ionisation balance. In the future, work will include the full modelling, which takes into account the high-lying levels, as well as other processes which influence the charge-state distribution in the TR to a lesser extent, such as auto-ionisation, photo-excitation, charge exchange and three-body recombination. The following section describes the sources used for the atomic rates incorporated into the model, and Section \ref{sec:crmmethod} discusses the steps taken to build the CR model itself.

\subsection{Atomic processes}
\label{sec:atomicmethod}

\subsubsection{Collisional ionisation}
\label{sec:collion}

Direct ionisation (DI) by electron impact is the main process, although, as suggested  by \cite{goldberg_etal:1965} for some isoelectronic sequences, additional, non-negligible ionisation can occur via inner-shell excitation into a state above the ionisation threshold which then spontaneously ionises (excitation--auto-ionisation, EA). This was confirmed with experiments, \citep[see e.g.][]{crandall1979}. 

Many theoretical and experimental studies of DI have been published over the years. The majority of studies have focussed on cross-sections from the ground level, except for Be-like ions, where ions in metastable levels are often present. For example, \cite{bell1983} presented a widely-used review of calculated and measured cross-sections between ground states of the main ions relevant for astrophysics, (it is noted that their equation 8 is incorrect). \cite{dere2007} used the Flexible Atomic Code \citep[FAC,][]{gu2008} to calculate DI and EA cross-sections for electron-impact ionisation, but only between ground states. Some of the calculated rates were also adjusted to agree with laboratory data. Together with the improved DR rates from the DR project, these rates formed a new, reference ionisation equilibrium, which was released for Chianti v.6 \citep{dere_etal:09_chianti_v6}.

Therefore, since rates are required for each level included in the model, it has been resolved to calculate all the DI cross-sections for the ground and excited states using FAC, as described below. The approach here differs from that of \cite{dere2007} in that the rates will not normally be adjusted to agree with experiment, especially since ions in excited states can be present, which affects the cross-sections measured. In fact, a large scatter (often beyond the quoted experimental uncertainties) in the laboratory measurements is present. This scatter and the differences with calculated values can be used as a measure of the uncertainty in the cross-sections, and will be used in a follow-up paper to provide uncertainties in the charge-state distribution. 

It has been noticed that, where the threshold for EA is close to that of DI, the EA rates can have an observable effect on ion populations. Consequently, level-resolved EA cross-sections have also been calculated for ground and excited states, this time using Autostructure \citep[AS,][]{badnell2011}. For each case, the calculations have been benchmarked with available theoretical and experimental studies.

For the modelling, the rate-coefficient\footnote{The rate-coefficients are being made available in electronic form} for a collisional process taking place from an initial level $j$ to a final level $i$ is related to the cross-section by:

\begin{equation}
C_{ij} = \int^{\infty}_{v_0} v~\sigma_{ij}(v)~f(v) \; dv \; ,
\end{equation}

\noindent where $v$ is the velocity of the impacting particle, $f(v)$ is the velocity distribution of these particles, the limits of the integral are from the ionisation-threshold velocity to infinity, and $\sigma_{ij}(v)$ is the cross-section for the process. The velocity distribution will be Maxwellian in thermal equilibrium. The rate-coefficient is multiplied by the number density of free particles involved in the collision to give the rate at which transitions  from $j$ to $i$ take place.

\subsubsection{Recombination}
\label{sec:rec}

For radiative recombination (RR), rates calculated by \cite{badnell2006} are used. For the DR rates at zero density, the values calculated by the DR project, led by N.R.~Badnell \citep[see][]{badnell2003}, are used. They are significantly different than the original rates used in the CR modelling by \citeauthor{burgess1969}.

Since this is the first stage in the modelling and high-lying levels have not been included, the total RR and DR rates from each of the ground and metastable levels of the recombining ion are used. Total recombination rates from an initial level in the recombining ion are the sum of recombination rates from that initial level to every possible level in the recombined ion. In this model, the total rates are applied to the ground level in the recombined ion. Although this may appear to overstate populations in the ground, collisional excitations and radiative decays are orders of magnitude faster than ionisation and recombination rates. Consequently, level populations in the recombined ions rapidly reach their equilibrium values before further ionisation or recombination takes place. Testing this method against using partial recombination rates into individual levels of the recombined ion shows in the case of RR from \ion{C}{iv} into \ion{C}{iii}, for example, a difference of 0.2\% in the ion populations.
	
In order to assess suppression of DR when density increases, \cite{nikolic2013} suggested recently an empirical formula to reproduce the DR suppression calculated by \cite{summers1974}. This is used to `correct' the new zero-density DR rates, in order to indicate when a density-dependent CR model is required. The main assumptions in this were that: a) the effective recombination rates provided by \cite{summers1974} were only due to the DR process; and, b) the behaviour with density would be the same as it would be calculated with the new DR rates. The approach does, however, apply only to total recombination rates and not to level-resolved rates. It is noted that \cite{young2018} used the DR project rates in combination with this approximate estimate of their suppression, using a corrected version of the \citeauthor{nikolic2013} approach, (which was published shortly afterwards \citep{nikolic2018}). 
	
The effect of density on DR has been approximated here by following a similar approach to \citet{nikolic2013, nikolic2018}. Unlike \citeauthor{nikolic2013}, here the DR suppression is estimated directly from the \cite{summers1974} rates, rather than relying on analytical formulae. This is carried out for every ion by calculating at each temperature point the ratio of the effective recombination rate at the particular density of interest to the effective recombination rate at the lowest density in the tables, $N_{\rm e} = 10^4$ cm$^{-3}$. The total DR rates of \citeauthor{badnell2003} at each temperature point are multiplied by this ratio to estimate the suppressed DR rate for the density in hand.

\subsubsection{Collisional excitation and radiative rates}
\label{sec:excrad}

For most ions, atomic data from Chianti v.8 \citep{delzanna_chianti_v8} for collisional excitation (by electron and proton impact) and radiative decay has been used. For \ion{C}{iii}, the improved collisional excitation calculated by \citet{fernandez2014} has been incorporated.

\subsubsection{Photo-ionisation}
\label{sec:pi}

Photo-ionisation can also be important in the lower TR, as shown by \citeauthor{nussbaumer1975}. Estimating the photo-ionising radiation is not a trivial issue. On-disc observations in the UV and EUV are strongly affected by solar variation. Irradiance measurements are in principle better but include limb-brightening effects. Variations of the radiances with the solar cycle can be significant, also \citep{andretta_delzanna:2014,delzanna_andretta:2015}.

The photo-ionising radiation used in the present model is derived from the quiet-Sun irradiances observed by the Solar Dynamics Observatory EVE instrument in the EUV, as reported by \citet{woods2009}. This set of data has a wide spectral range, which covers the appropriate wavelengths for all ion stages of carbon. The irradiances were cross-checked with the higher-resolution radiances given by \citet{curdt2001}, as observed by the Solar Heliospheric Observatory SUMER instrument. Its narrow range meant it was not suitable to be used for the rates, but comparisons showed good overall agreement with the radiances derived from \citeauthor{woods2009}

The solar radiation adopted by \citeauthor{nussbaumer1975} was taken from a variety of sources. The main source used in the work was \cite{malinovsky_etal:1973} for the EUV. It is noted that these are irradiance observations obtained when the Sun was very active \citep[see][]{delzanna2019}; it is not straightforward to convert irradiances to radiances in this case. In fact, even for normal, quiet-Sun conditions different spectral lines have a different limb-brightening, as discussed by \cite{andretta_delzanna:2014}. In some cases, radiances turned out to be incorrect by factors of about 2, \citep[see][and references therein]{delzanna_andretta:2015}.

The analytical fits to the Opacity Project, R-Matrix cross-section backgrounds derived by \citet{verner1996}, widely used in the literature for modelling, are not suitable for the level-resolved picture since they give total rates only from the ground level. In this CR model it would not take account of photo-ionisation of ions in excited levels, nor the differing ionisation rates of those levels. Consequently, level-resolved cross-sections are used, as calculated by \citet{badnell2006} and made available on the APAP Network website\footnote{www.apap-network.org}. These were checked by comparing the total rates of the ground level with the \citeauthor{verner1996} fits, and were found to be in very good agreement.

The matrix elements for photo-ionisation from a bound level $j$ to a final level $i$ are given by:

\beq
\alpha^{PI}_{ij} = 4 \pi \; \int\limits_{\nu_0}^\infty {\sigma_{ij}(\nu) \over h\nu} \; J_\nu \; {\rm d}\nu \; ,
\eeq

\noindent where $\nu_0$ is the threshold frequency below which the bound-free cross-section $\sigma_{ij}(\nu)$ is zero. 

\beq
J_{\nu} = { \Delta\;  \Omega  \over 4 \pi} \;  \overline{I_\nu} = 
W(r) \; \overline{I_\nu} \; ,
\eeq

\noindent where $W(r)$ is the {\it dilution factor} of the radiation, that is, the geometrical factor  which accounts for the weakening of the radiation field at a distance $r$ from the  Sun, and $\overline{I_\nu}$ is the averaged disc radiance at frequency $\nu$.

\subsection{Collisional-radiative modelling}
\label{sec:crmmethod}

The population $N^z_{i}$ (for each temperature) of an ion with charge $z+$ in a level $i$ can be obtained from:

\begin{equation}
\eqalign{
	\frac{dN^{z}_{i}}{dt} = & \sum_{j}C^{z}_{ij}N^{z}_{j} + \sum_{j}S^{z-1}_{ij}N^{z-1}_{j} + \sum_{j}R^{z+1}_{ij}N^{z+1}_{j} \cr
	& - \sum_{j}C^{z}_{ji}N^{z}_{i} - \sum_{j}S^{z}_{ji}N^{z}_{i} - \sum_{j}R^{z}_{ji}N^{z}_{i} ~, \cr}
\end{equation}

\noindent where $C^{z}_{ij}$ represents the collisional-radiative matrix element for processes within an ion from level \textit{j} to level \textit{i}, $S^{z}_{ij}$ is the matrix element for ionisation processes from level \textit{j} of charge-state $z+$ into level \textit{i} of the next higher charge-state, and $R^{z}_{ij}$ is the element for recombination processes out of level \textit{j} in charge-state $z+$ into level \textit{i} of the next lower charge-state. In addition, there is the normalisation condition that the total ion populations should be equal to the elemental abundance: $N(X)=\sum_z N^{z}$. In ionisation equilibrium, ${dN^z_i \over dt} = 0$.

The usual method to calculate ion charge-states at zero density (as in e.g. the Chianti database) is to consider the total ionisation and recombination rates between successive charge-states considering only the ground states. The relative population of two ions is then obtained directly from the ratio of the total ionisation and recombination rates at each temperature. 

To take into account density effects, a level-resolved model has been developed in which all the main fine-structure levels for the carbon ions have been included. Matrices which include all the main rates were built for each ion. The matrix elements for the processes occurring within one ion were obtained in the same way as in the Chianti package. They include collisional excitation and de-excitation by electron and proton impact, photo-excitation and -de-excitation, plus radiative decay. The other elements of the matrices, for transitions between ions, were populated using the ionisation and recombination rates for the ground and metastable levels which were described above. The populations of all the levels for all ions are then solved at once. This is a novel approach; it uses a significant modification of the Chianti codes, and has been written in IDL. Part of these codes have been included in v.9 of Chianti \citep{dere2019}, where a matrix for two ions at once is solved, to calculate the intensities of the satellite lines.

\section{Results}
\label{sec:results}

\subsection{Atomic data}
\label{sec:atomicresults}

\subsubsection{Direct ionisation}
\label{sec:ionresults}

Level-resolved  DI cross-sections for each charge-state of carbon, up to \ion{C}{v}, were computed using the semi-relativistic, distorted-wave (DW) method as implemented in FAC. Comparisons were made with available experiments and theoretical calculations in order to validate the results. Experiments where ions in metastable levels are present provide a valuable way of validating cross-sections for both ground and metastable levels at the same time, so long as the populations of ions in metastable levels are known to a reasonable degree of accuracy.

To model the \ion{C}{ii} populations effectively, \ion{C}{i} ionisation rates are required. Relatively recently, \citet{abdel2013} carried out theoretical cross-section computations, utilising R-Matrix, time-dependent close-coupling (TDCC) and DW methods. However, they did not give a sufficiently high energy range to be able to use the results for ionisation rates. Calculations of ionisation cross-sections were made with FAC, whilst bearing in mind the known shortcomings of the DW approximation for near-neutral charge-states. The main experiment used for comparison has been \citet{brook1978}, which \citet{bell1983} and \citet{Suno2007} use for their recommended data. 

\begin{figure}
	\centering
	\includegraphics[width=9.8cm]{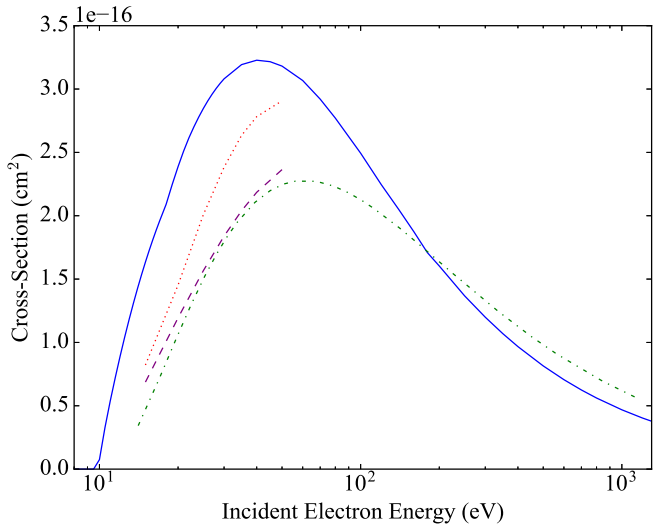}
	\caption[width=0.5\columnwidth]{Configuration-average direct ionisation cross-section for \ion{C}{i} ground state; blue solid line - this work, green dash-dotted - \citeauthor{bell1983}, purple dashed - \citeauthor{abdel2013} TDCC, red dotted - \citeauthor{abdel2013} DW.}
	\label{fig:c1grd}
\end{figure}

For the ground state, the non-perturbative R-Matrix and TDCC calculations of \citeauthor{abdel2013} lie very close to the experimental data. Their DW result lies 18\% above the other results at the peak. For the same level, the FAC cross-section is almost 30\% above the recommended data, as shown in Figure \ref{fig:c1grd}. For the first excited configuration $2s~2p^3$, the DW calculations from FAC and \citeauthor{abdel2013} lie 39\% and 24\% above TDCC values, respectively. For ionisation from the $2s~2p^2~3l$ configurations, FAC is equal to the TDCC calculations, while their DW results diverge.

For rate-coefficients, accuracy in the near-threshold behaviour of the cross-section is important, which can be another shortcoming of the DW approach. A simple reduction of the FAC cross-sections by 30\% does not resolve the situation. To solve the problem and provide rates for the model, the recommended data of \citeauthor{bell1983} has been adjusted in height, based on the ratio of the FAC excited level cross-sections to the FAC ground level cross-section. The scaled energies of \citeauthor{bell1983} were converted to incident energies by using the FAC ionisation-threshold energies for each level, which were in good agreement with \citeauthor{abdel2013}

\begin{figure}
	\includegraphics[width=9.8cm]{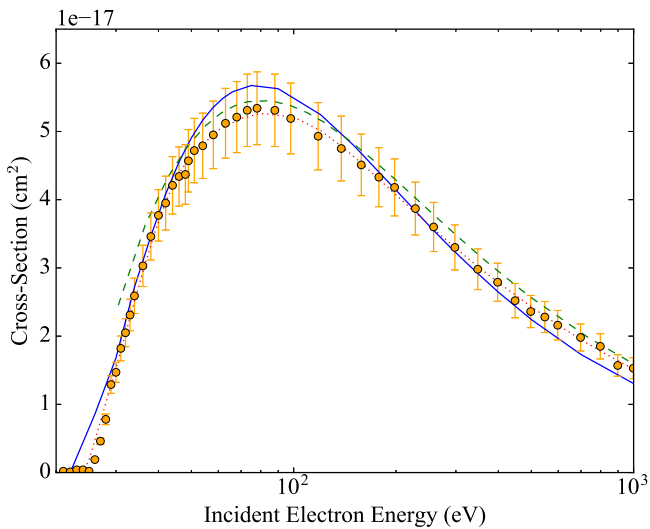}
	\caption[width=1.0\columnwidth]{Direct ionisation cross-section from \ion{C}{ii} ground level; solid blue line - this work, red dotted - Chianti v.8, green dashed - \citeauthor{bell1983}, orange circles - \citeauthor{yamada1989}.}
	\label{fig:c2grd}
\end{figure}

For \ion{C}{ii}, \citet{bell1983} follows the experiment of \citet{aitken1971} for recommended values because \citet{hamdan1978} has noticeable below-threshold values. A later experiment by \citet{yamada1989} has no below-threshold behaviour and lies a few per cent below \citeauthor{aitken1971} Figure \ref{fig:c2grd} shows that the FAC ground level cross-section follows a very similar energy distribution and peaks just a few per cent above \citeauthor{bell1983} Chianti uses the ground-level cross-section of \citet{dere2007}, who also used FAC. The R-Matrix result of \citet{ludlow2008} is very close to the \citeauthor{yamada1989} experiment. The only available comparison for the metastable term is with the work of \citeauthor{ludlow2008} Their configuration-average, DW cross-section is 5\% higher than this work, but their term-resolved, R-Matrix cross-section is 25\% higher, although the R-Matrix result includes EA, (see Section \ref{sec:earesults}). The rate-coefficients in Chianti for the ground level are 10\% lower than this work.

\begin{figure}
	\centering
	\includegraphics[width=9.8cm]{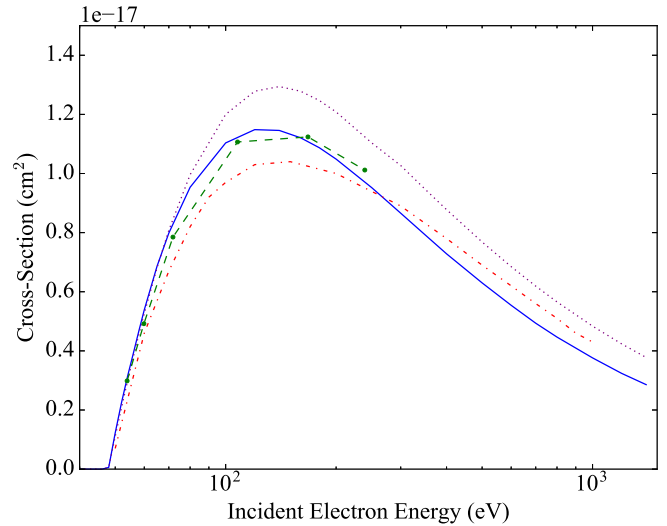}
	\caption[width=1.0\linewidth]{Direct ionisation cross-section for \ion{C}{iii} ground level; blue solid line - this work, green dashed - \citeauthor{younger1981be}, red dash-dotted - \citeauthor{fogle2008} R-Matrix, purple dotted - Chianti v.8.}
	\label{fig:c3grd}
\end{figure}

\begin{figure}
	\centering
	\includegraphics[width=9.8cm]{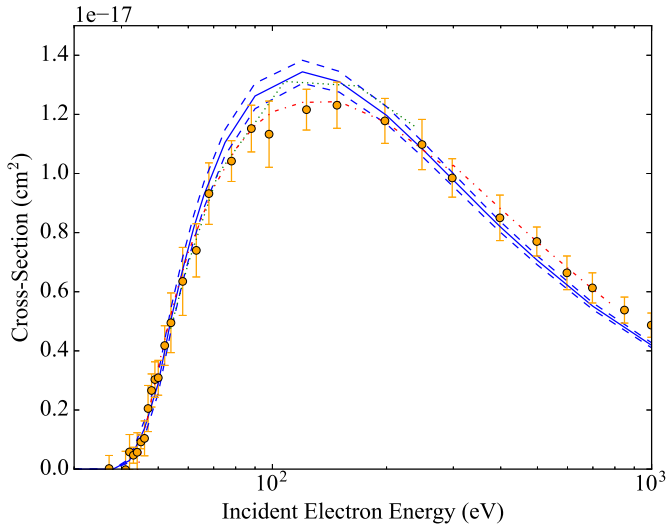}
	\caption[width=1.0\columnwidth]{\small Combined direct ionisation cross-section for \ion{C}{iii} with 40\% metastable population; blue solid line - this work, blue dashed - difference in cross-section owing to metastable population uncertainty, green dotted - \citeauthor{younger1981be}, red dash-dotted - \citeauthor{fogle2008} R-Matrix, orange circles - \citeauthor{woodruff1978} experiment.}
	\label{fig:c3expt}
\end{figure}

For \ion{C}{iii}, the FAC cross-section for the ground state lies close to the DW work of \citet{younger1981be} and 10\% above the R-Matrix values of \citet{fogle2008}, which are shown in Figure \ref{fig:c3grd}. Chianti is a further 10\% higher. The cross-section of the excited term $2s~2p~^3P$ in the current work is very close to the earlier DW calculation of \citeauthor{younger1981be} and less than 10\% higher than the R-Matrix approach. The \citeauthor{fogle2008} experiment was able to determine the proportion of metastable ions in the beam more accurately, but they have fewer energy points and a larger scatter than \citet{woodruff1978}. \citeauthor{woodruff1978} also has more energy points close to threshold, which is useful for comparing with the EA calculations discussed in Section \ref{sec:earesults}. \citet{falk1983} estimates the \citeauthor{woodruff1978} metastable fraction to be 40\% and comparison with the R-Matrix cross-sections combined according to these populations suggests this is reasonable (Figure \ref{fig:c3expt}). An estimated uncertainty of 20\% in the metastable population (c.f. the 15\% uncertainty in the \citeauthor{fogle2008} experiment) is shown by the dashed lines either side of the combined DI cross-sections from this work. It is seen that the theoretical cross-sections are almost entirely within the experimental uncertainties. The rate-coefficients calculated using the FAC cross-sections and a Maxwellian electron distribution are all within 10\% of the \citeauthor{fogle2008} R-Matrix rate-coefficients for this ion.

\citeauthor{bell1983} considered the \ion{C}{iv} values of the \citet{crandall1979} experiment to be too low and recommended the theoretical results of \citet{jakubowicz1981} Coulomb-Born-Exchange approximation. \citet{dere2007} is close to the \citet{knopp2001} experimental values. Calculations from FAC lie less than 10\% below \citeauthor{bell1983} Similarly, for \ion{C}{v} the ground state cross-sections are close to \citet{bell1983}, who believed the \citet{crandall1979} experimental technique produced inherently high values for this ion. \citet{dere2007} matches \citeauthor{crandall1979}

\subsubsection{Excitation--auto-ionisation data}
\label{sec:earesults}

To calculate the cross-sections for EA requires data from several processes: collisional excitation, auto-ionisation and radiative decay. A bound electron is collisionally excited to an auto-ionising state above the threshold, and then it may either spontaneously ionise or decay. Calculating EA cross-sections involves multiplying the excitation cross-section to a level by the auto-ionisation branching ratio from that level, and summing over all levels which lie above the ionisation threshold. The auto-ionisation branching ratio is the probability that auto-ionisation takes place instead of decay. The EA cross-section for an ion in an initial level $j$ to ionise to a final level $i$ in the next higher charge-state, through a collision with a free electron with energy $\epsilon$, is given by:

\begin{equation}
	\sigma^{~EA}_{ij}(\epsilon) = \sum_p \; \sigma_{pj}(\epsilon)~ \frac{A^a_{ip}}{\sum_q A^a_{qp} ~+~ \sum_s A^r_{sp}}~,
\end{equation}

\noindent where $\sigma_{pj}(\epsilon)$ is the electron-impact excitation cross-section to an auto-ionising level $p$, $A^a$ is the auto-ionisation rate and $A^r$ is the radiative decay rate. The sum over $p$ is for excitations to all levels within the ion over the ionisation limit, the sum over $q$ is for all possible final levels in the next higher charge-state to which the ion may spontaneously ionise, and the sum over $s$ is for all possible levels to which radiative decay may take place within the existing ion. Autostructure was used to obtain data for the required processes, which were then combined to create level-resolved rate-coefficients, which are available online. Only excitations to energy levels shown by NIST\footnote{physics.nist.gov/asd} \citep{kramida2018} to be above the ionisation limit were included.

\begin{figure}
	\centering
	\includegraphics[width=9.8cm]{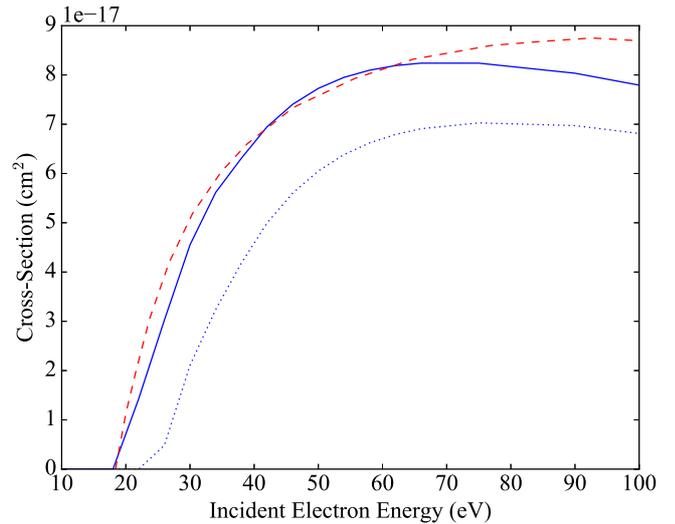}
	\caption[width=1.0\linewidth]{Total collisonal ionisation cross-section for \ion{C}{ii} metastable term, $2s~2p^2 \; ^4\rm{P}$; blue solid line - this work, blue dotted - this work (DI only), red dashed - \citeauthor{ludlow2008} R-Matrix.}
	\label{fig:c2metatotal}
\end{figure}

As mentioned in Section \ref{sec:ionresults}, the DW method is less reliable for neutral and near-neutral charge-states. This was observed when computing EA cross-sections for \ion{C}{ii}, which showed variations of factors of two to three in the cross-section depending on the structure of the ion. To resolve this, the structure used by \citet{liang2012} for carbon was utilised. This produced a total EA cross-section for the ground term, $2s^2~2p \; ^2\rm{P}$, which has a peak of $1.2 \times 10^{-17}$ cm$^{-3}$ at 55~eV, (c.f. the DI cross-section in Fig. \ref{fig:c2grd}). At the same energy, the \citet{ludlow2008} DW cross-section is 20\% higher. Their R-Matrix cross-sections include both DI and EA. Compared to those results, the total DI and EA cross-section from this work is 26\% higher at the peak. The conclusion of \citeauthor{ludlow2008} noted how electron coupling effects, which are inherently included in R-Matrix calculations, can reduce total cross-sections by 15-35\%. For the metastable term, $2s~2p^2 \; ^4\rm{P}$, the total DI and EA cross-section at the peak is 8\% below their R-Matrix cross-section (Figure \ref{fig:c2metatotal}).

The uncertainty observed in the cross-sections, which arose from differences in structure, was also seen in the rate-coefficients. Because of this variability, the EA rate-coefficients were computed using the R-Matrix, thermally-averaged collision-strengths from \citeauthor{liang2012} in combination with AS auto-ionisation data, which was calculated using an identical structure. Since the R-Matrix collision-strengths only include outer-shell excitations, inner-shell EA has been added to the rate-coefficients by using AS DW excitation and auto-ionisation data with the same structure. Inner-shell EA contributes significantly less to the total EA over most of the temperature range for this ion. As an example of the contribution made by EA for this ion, at 40,000~K the ground level, EA rate-coefficient is one-third of the ground level, DI value. Because of the uncertainty found with the DW approximation for \ion{C}{ii} and the absence of similar R-Matrix excitation data for \ion{C}{i}, no attempt has been made to produce EA rate-coefficients for \ion{C}{i} at this stage.

\begin{figure}
	\centering
	\includegraphics[width=9.8cm]{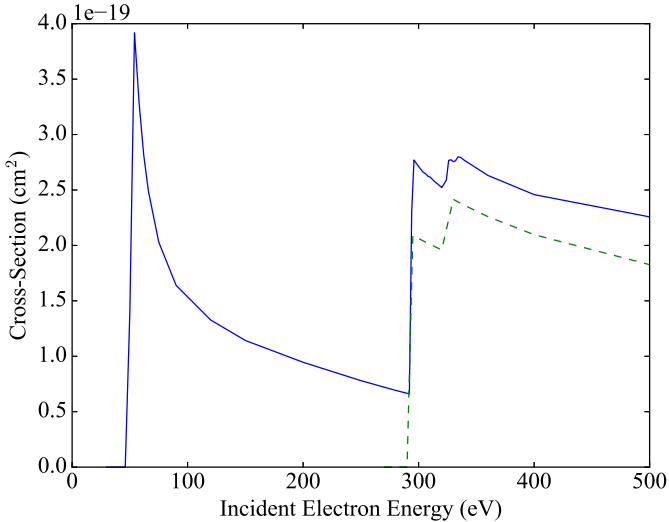}
	\caption[width=1.0\linewidth]{Total EA cross-section for \ion{C}{iii} ground level; blue solid line - this work, green dashed - Chianti v.8.}
	\label{fig:c3ea}
\end{figure}

For \ion{C}{iii}, EA from the ground and metastable levels is possible through excitations of the $2s^2$ and $2s~2p$ electrons to the $2p~4d~^3\rm{D}$ term and above followed by auto-ionisation to the ground level of \ion{C}{iv} \citep[c.f.][]{loch2005}. This is seen in the cross-section shown in Figure \ref{fig:c3ea}. The inner-shell cross-section of this work is within 5\% of Chianti, which does not have the outer-shell excitations. The main contributions for the ground level come from weaker two-electron transitions, whereas for the metastable $2s~2p~^3\rm{P}$ term the dominant contributions come from one-electron excitations. Consequently, the metastable-term EA cross-section is an order of magnitude higher, which is in agreement with the value given by \citet{loch2005}, that is, about 10\% of the DI value. For EA from ground and metastable levels to final levels in \ion{C}{iv} above the ground level the only contribution to EA is from K-shell excitations. 

Cross-sections were much less prone to variations caused by the structure than for \ion{C}{ii}; differences in structure altered the results by 10\% at the most. However, owing to sensitivity of the collision-strengths to the mixing coefficients and weak, two-electron excitations from the ground, EA rate-coefficients were affected by which approximation was used for the scattering computation. Consequently, EA rate-coefficients were computed using R-Matrix collision-strengths and AS auto-ionisation data, in the same way as for \ion{C}{ii}. This time, the excitation data was from \citet{fernandez2014}. Using the R-Matrix data also has the advantage of including the resonance contributions. Again, since only L-shell excitations are available for \ion{C}{iii} in \citeauthor{fernandez2014}, to these were added unitarised DW, K-shell contributions, employing the same structure as the R-Matrix data. Inner-shell EA becomes noticeable in the rate-coefficients for temperatures above $10^6$ K when calculated with a Maxwellian electron distribution.
	
\begin{figure}
	\centering
	\includegraphics[width=9.8cm]{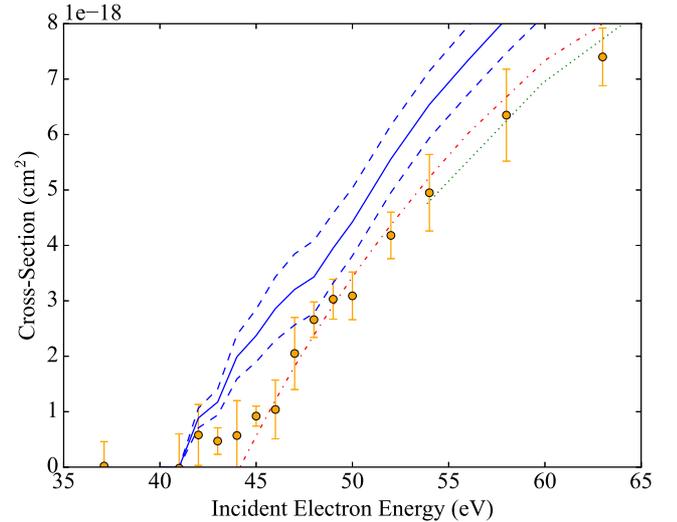}
	\caption[width=1.0\linewidth]{Combined total ionisation cross-section for \ion{C}{iii} with 40\% metastable population; blue solid line - this work, blue dashed - difference in cross-section owing to metastable population uncertainty, green dotted - \citeauthor{younger1981be}, red dash-dotted - \citeauthor{fogle2008} R-Matrix, orange circles - \citeauthor{woodruff1978} experiment.}
	\label{fig:c3totalioniz}
\end{figure}

A comparison close to the threshold of the total ionisation cross-section of \ion{C}{iii} with the experiment of \citeauthor{woodruff1978} is shown in Figure \ref{fig:c3totalioniz}. It may be possible to observe some peaks close to the relevant thresholds, but, given that the combined uncertainties in the metastable population and cross-section measurement is greater than the EA contribution, these may simply be scatter in the results.

\begin{figure}
	\centering
	\includegraphics[width=9.8cm]{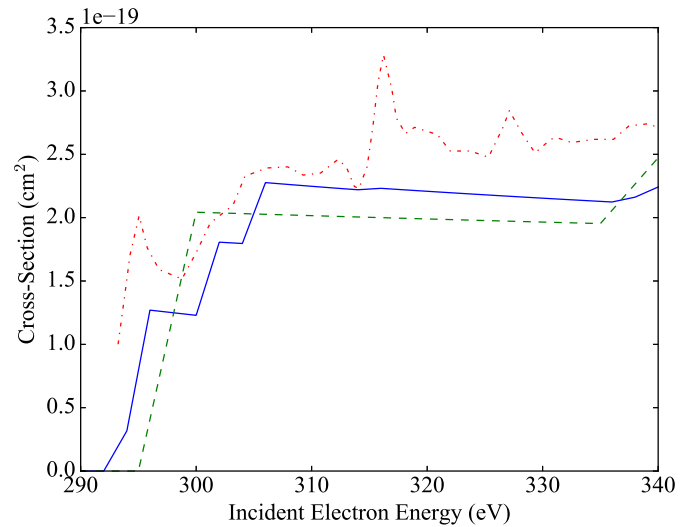}
	\caption[width=1.0\linewidth]{Total EA cross-section for \ion{C}{iv} ground level; blue solid line - this work, green dashed - Chianti v.8, red dash-dotted - \citeauthor{teng2000} experiment.}
	\label{fig:c4ea}
\end{figure}

For \ion{C}{iv} inner-shell excitations provide the only possibility for EA to take place. The cross-section for the ground level is within 10\% of Chianti. \citet{teng2000} estimated the uncertainty in their experimental values to be 20-30\%, which encompasses both of the cross-sections shown here (Figure \ref{fig:c4ea}). The DW rate-coefficients agree with the R-Matrix rate-coefficients, produced from the thermally-averaged collision-strengths calculated by \citet{liang2011}, to within 5\%, but the R-Matrix ones are made available here for the sake of completeness.

\subsubsection{Multiple-ionisation cross-sections}

\citet{hahn2017} compiled a set of experimental, multiple-ionisation cross-sections for a wide range of astrophysically relevant ions. From these, they provided geometrical fitting formul{\ae} to reproduce the cross-sections. The formul{\ae} they provided for cross-sections have been used to calculate rates, which were included in the CR model. 

Double, direct ionisation produces cross-sections which are between 0.5\% and 4\% of single DI values. Ionisation followed by auto-ionisation is about an order of magnitude lower than double DI, except in the case of \ion{C}{ii}, where it is comparable, but at an higher threshold. Triple DI is two orders of magnitude less than double DI, and has not been included in the modelling. Another factor to bear in mind for the CR model is that the threshold for double ionisation is much higher, and so, compared to single ionisation, the rates will be significantly lower. The rates may be more significant with non-Maxwellian electrons distributions which have a strong high-energy component.

\subsection{Collisional-radiative model}
\label{sec:crmresults}

As described in Section \ref{sec:crmmethod}, all rates available in Chianti except collisional ionisation were imported into the model, followed by the collisional- and photo-ionisation rates calculated in this work. The effect of each of the processes is determined first by running the model in the coronal approximation, that is, a single, ground-level ion with total rates between neighbouring charge-states. This assists identifying the principal effect each process has on the charge-state distribution. Thereafter, the model is switched to level-resolved modelling for each of the atomic processes to determine the full effect on the ion populations. From Chianti, data is imported from 42 levels of \ion{C}{i}, 204 levels of \ion{C}{ii}, 238 of \ion{C}{iii} and 331 of \ion{C}{iv}. The remaining charge-states are each modelled as a single level. These ion models are valid for solar densities, up to flare densities of $N_{\rm e}=10^{13}$ cm$^{-3}$.

\subsubsection{CR model with collisional ionisation from metastables}
\label{sec:crmion}

\begin{figure}
	\centering
	\includegraphics[width=9.4cm]{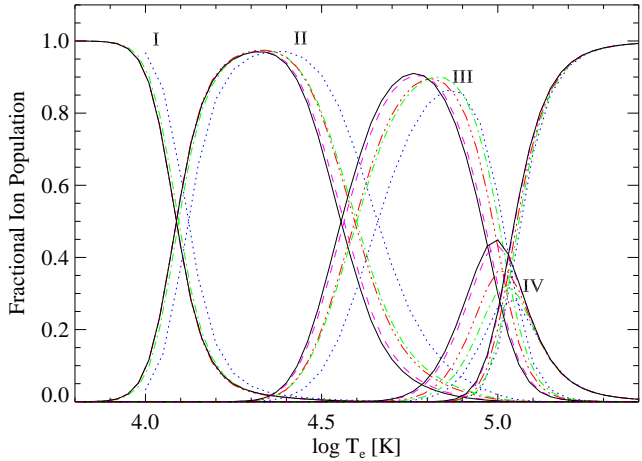}
	\caption{Effect with density of level-resolved, collisional ionisation on the CR model; blue dotted line - Chianti v.8, green dash-dotted - this work at $10^4$ cm$^{-3}$ density, red dash-dot-dotted - $10^8$ cm$^{-3}$, purple dashed - $10^{10}$ cm$^{-3}$, black solid - $10^{12}$ cm$^{-3}$.}
	\label{fig:crmlrdens}
\end{figure}

It is at densities which redistribute ion populations to terms above the ground term that produce the first noticeable effects of shifting the temperature formation of ions in the level-resolved model, compared to the single-level model. For \ion{C}{i}, the term above the ground is populated at a density of $10^4$ cm$^{-3}$. Therefore, the difference in \ion{C}{i} populations at this density between the single-level model, which is used by Chianti, and the level-resolved model of this work is owing to ionisation occurring from the metastable levels (Figure \ref{fig:crmlrdens}). For \ion{C}{ii} and \ion{C}{iii}, however, the term above the ground is not appreciably populated until the density reaches $10^8-10^{10}$ cm$^{-3}$. Consequently, the difference between the populations of the single-level and level-resolved models for those ions at densities lower than $10^8$ cm$^{-3}$ arises from the differing ionisation rates for the ground term being used by each model. At $10^{10}-10^{12}$ cm$^{-3}$ the metastable populations have reached statistical balance in these ions, and so, above these densities, the effect collisional ionisation plays through metastable levels reaches saturation. To illustrate these effects, at 40,000K the total \ion{C}{ii} population is 47\% with a density of $10^4$ cm$^{-3}$, 45\% at $10^8$ cm$^{-3}$, 34\% at $10^{10}$ cm$^{-3}$ and 31\% at $10^{12}$ cm$^{-3}$.

The changes in the ionisation balance also affect the temperature at which ions peak in their abundance. At a density of $10^{12}$ cm$^{-3}$ \ion{C}{ii} now forms its peak at 21,000K, instead of 24,000K at zero density; for \ion{C}{iii} it is now at 56,000K from 76,000K, and forms 92\% of the population at the peak, instead of 87\%. \ion{C}{iv} witnesses a 55\% increase in abundance at its peak, from a fractional population of 0.29 to 0.45. Since it has no metastable levels, the ionisation rates of \ion{C}{iv} do not change with density, and so the increased populations seen in both \ion{C}{iv} and \ion{C}{v} are attributable to the formation of \ion{C}{iii} at lower temperatures in the level-resolved model.

\begin{figure}
	\centering
	\includegraphics[width=9.4cm]{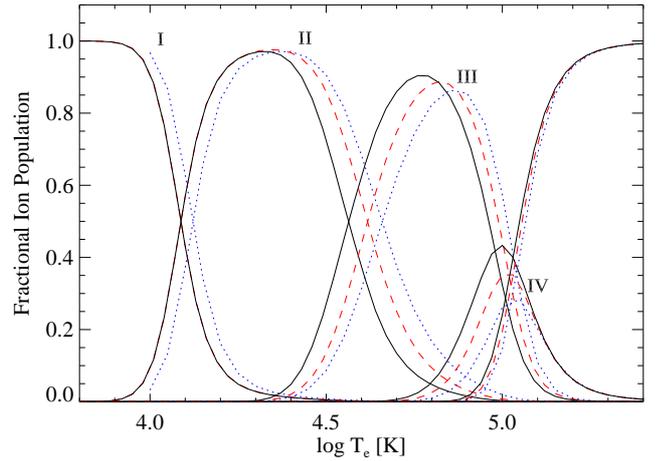}
	\caption{Changes in the effect of level-resolved, collisional ionisation on the CR model at $10^{10}$ cm$^{-3}$ density when EA is excluded; black solid line - this work with EA, red dashed - this work without EA, blue dotted - Chianti v.8.}
	\label{fig:crmlrdensea}
\end{figure}

Many CR models only include K-shell EA, \citep[see e.g.][]{arnaud_rothenflug:85,avrett2008}. However, the threshold energy is far above the formation temperatures of the ions, and makes negligible changes to the ion populations. It is outer-shell EA that is making an equal contribution to DI in the shift in ion formation (Figure \ref{fig:crmlrdensea}). \citet{goldberg_etal:1965} indicated that EA would make an important contribution to ion populations for B-like \ion{O}{iv}, but not for Be-like \ion{O}{v}. For carbon, EA is seen to be important for both B-like and Be-like ions. Since the ionisation potential of \ion{C}{i} is small, there would be many excited configurations which would contribute to EA of this ion, and so it is reasonable to expect the formation temperature of this ion would be lowered further. This would broaden the temperature range over which \ion{C}{ii} forms, increasing its contribution to emission. Multiple ionisation has no noticeable influence on the charge-state distribution in these conditions.

\subsubsection{CR Model with DR suppression}
\label{sec:crmdr}

\begin{figure}
	\centering
	\includegraphics[width=9.4cm]{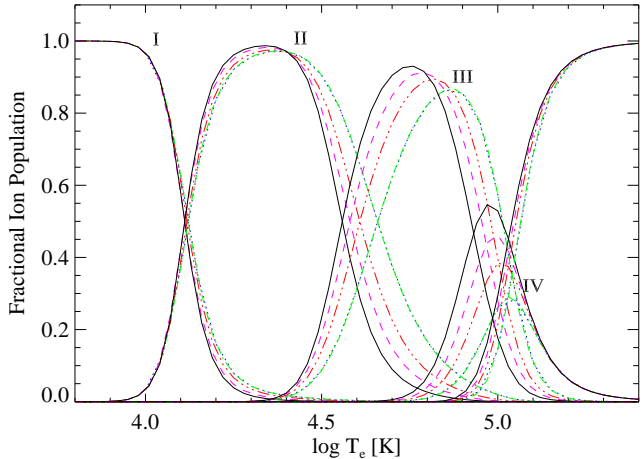}
	\caption{Effect with density of DR suppression on the ground-to-ground CR model; blue dotted line - Chianti v.8, green dash-dotted - this work at $10^4$ cm$^{-3}$ density, red dash-dot-dotted - $10^8$ cm$^{-3}$, purple dashed - $10^{10}$ cm$^{-3}$, black solid - $10^{12}$ cm$^{-3}$.}
	\label{fig:crmdrsuppr}
\end{figure}

Although DR suppression is only approximated here, it gives clues to the behaviour seen in full CR models, in which high $n$ levels are included. Looking at the impact of DR suppression on the single-level, coronal-approximation model first, in order to separate this phenomenon from the effect of collisional ionisation, the only noticeable effect on \ion{C}{i} is in the reduction of populations at the higher temperature end of its formation, above log~$T_{\rm e}$(K)~=~4.1 (Figure \ref{fig:crmdrsuppr}). This can be accounted for in the DR rates of \ion{C}{ii}, which are at their weakest in the range 2,000-15,000K. Suppressing these rates has little impact on the charge-state distribution. 

The DR rates for \ion{C}{iii} are strong in the region of log~$T_{\rm e}$(K)~=~4.6--5.9, explaining the greater effect between the populations of \ion{C}{ii} and \ion{C}{iii} at temperatures above log~$T_{\rm e}$(K)~=~4.5. The suppression causes a shift in the peak abundance of \ion{C}{iii} from 76,000K to 55,000K and the peak population from 0.87 to 0.93 at a density of $10^{12}$ cm$^{-3}$ compared to zero density, which is similar to the effect level-resolved, collisional ionisation has on the model.

The consequent effect of lowering the formation temperature of \ion{C}{iii} is, again, to substantially increase abundance of \ion{C}{iv}. This time, the peak abundance of \ion{C}{iv} increases from 0.29 at density $10^4$ cm$^{-3}$ to 0.56 at $10^{12}$~cm$^{-3}$, a rise of 93\%, more significant than the rise seen with collisional ionisation. 

The behaviour seen in \ion{C}{iv} through density suppression of DR is explained by two factors. Firstly, the DR rate from \ion{C}{iv} is much larger than radiative recombination at its formation temperature. Thus, suppressing the DR rate causes more ions to remain in this ionisation stage, and reducing the populations in the stage below. Secondly, it is explained by a relatively small change in abundance of \ion{C}{v}. Physically, this is based on the closed-shell nature of \ion{C}{v}. The ions will be almost completely in the ground level, which requires large free-electron energies to excite bound, K-shell electrons to the next, L-shell level in order for DR to occur. Thus, the \ion{C}{v} DR rates are not significant until temperatures rise above $10^6$~K, far higher than where \ion{C}{iv} forms. The ions are being formed by the interplay of collisional ionisation of \ion{C}{iv} and radiative recombination from \ion{C}{v}, processes which have significantly lower thresholds. Consequently, suppressing DR has little effect in reducing recombinations from \ion{C}{v} into \ion{C}{iv}.

\subsubsection{CR Model with both metastable ionisation and DR suppression }
\label{sec:crmlrdr}

\begin{figure}
	\centering
	\includegraphics[width=9.4cm]{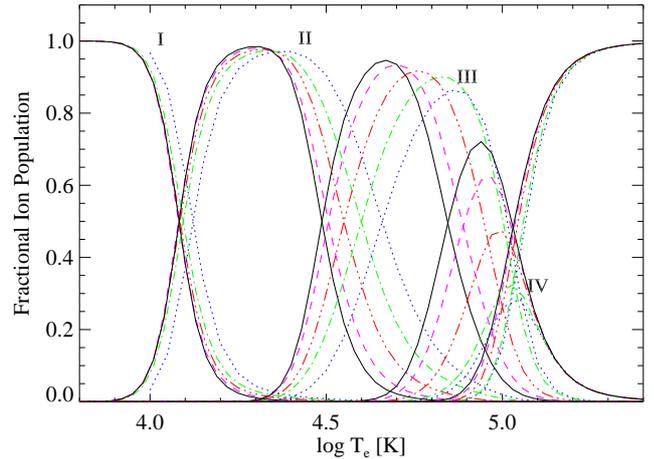}
	\caption{Combined effect with density of level-resolved, collisional ionisation and DR suppression on the CR model; blue dotted line - Chianti v.8, green dash-dotted - this work at $10^4$ cm$^{-3}$ density, red dash-dot-dotted - $10^8$ cm$^{-3}$, purple dashed - $10^{10}$ cm$^{-3}$, black solid - $10^{12}$ cm$^{-3}$.}
	\label{fig:crmlrdr}
\end{figure}

Taking density effects on the model through the combination of level-resolved collisional ionisation and dielectronic recombination suppression, (shown in Figure \ref{fig:crmlrdr},) the change between \ion{C}{i} and \ion{C}{ii} comes through ionisation of the metastable levels in \ion{C}{i}. Between \ion{C}{ii} and \ion{C}{iii}, the individual effects of collisional ionisation and DR suppression are comparable and both contribute to lowering the peak formation temperatures. Level-resolved ionisation from \ion{C}{iii} has a lesser impact on the population of \ion{C}{iv}, and so it is the DR suppression of \ion{C}{iv} that is influencing its increase in population the most. Since there are no metastable levels in \ion{C}{iv}, the only increase in formation of \ion{C}{v} at a lower temperature is owing to the increased population of \ion{C}{iv}. 

Looking at the charge-state distribution as a whole, the reduction in temperature at which \ion{C}{i} ionises combined with almost no change in the formation temperature of \ion{C}{v} means there is a wider range over which the intervening ions may form. There is a consequent shift towards lower temperatures and higher peak abundances for those ions.

\subsubsection{Photo-ionisation effect on ion populations}
\label{sec:crmpi}

\begin{figure}
	\centering
	\includegraphics[width=9.4cm]{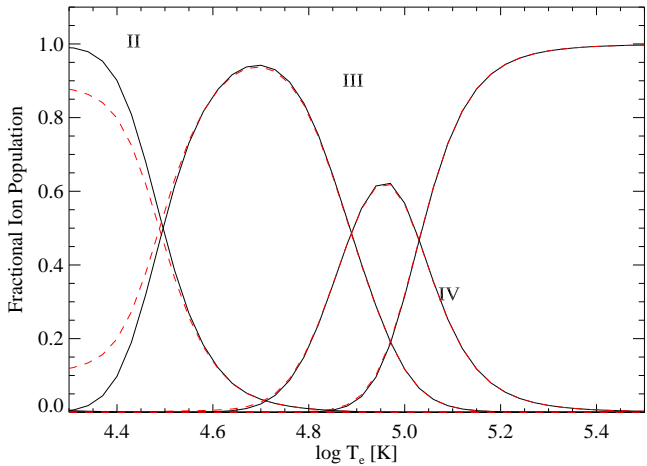}
	\caption{Effect of photo-ionisation on the level-resolved CR model at $10^{10}$ cm$^{-3}$ density; black solid line - this work without PI, red dashed - this work with PI.}
	\label{fig:crmlrdrpi}
\end{figure}

PI rates become weaker with increasing charge-state, such that \ion{C}{v} and \ion{C}{vi} are unaffected by this process; collisional processes are dominating the higher density plasma conditions typical of the solar and stellar atmospheres which are being modelled here. Since the PI rates are independent of temperature and density, in conditions of higher density the collisional rates dominate. So, based on the quiet-Sun radiances used in this model, at densities above $10^{12}$ cm$^{-3}$ the charge-state distributions when PI is included are the same as those without PI. Another consequence of this aspect of the PI rates is that at lower densities, such as TR densities of around $10^{10}$ cm$^{-3}$, the usual formation curves of \ion{C}{iii} and \ion{C}{iv} exist, but a small population of \ion{C}{iii} (12\%) is present at lower temperatures, arising from PI of \ion{C}{ii} (Figure \ref{fig:crmlrdrpi}). 

Since the PI rates depend on the particular radiances present for each of the transitions, it is difficult to determine systematic differences when switching between the coronal approximation and the level-resolved model. In the particular scenario simulated here, in the ground-to-ground model where total rates are used for all the processes, the \ion{C}{iii} population is 7\% at 20,000~K, instead of 12\%. The scenario demonstrates, as with collisional ionisation, level-resolved photo-ionisation produces greater populations in the next higher charge-states because of the faster ionisation rates from metastable levels.

Stronger radiances in other conditions, when active regions are present or during flares, for instance, should result in greater population shifts towards higher charge-states.

\subsubsection{Comparisons with other CR models}
\label{sec:crmnuss}

\begin{figure}
	\centering
	\includegraphics[width=9.4cm]{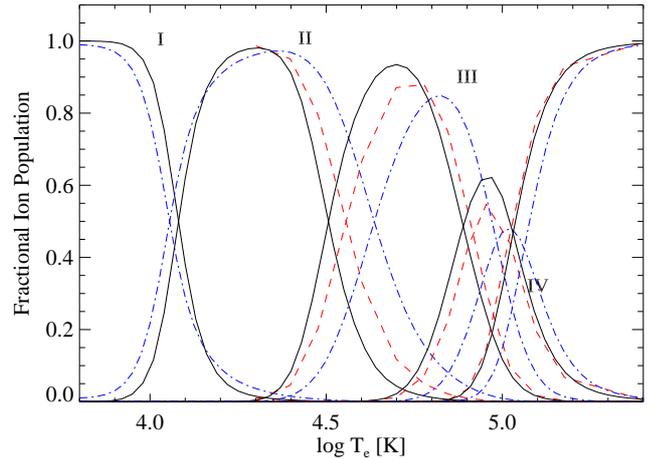}
	\caption{Comparison of the CR model with other works at $10^{10}$ cm$^{-3}$ density; black solid line - this work, red dashed - \citeauthor{nussbaumer1975}, blue dash-dotted - \citeauthor{summers1974}.}
	\label{fig:crmnuss}
\end{figure}

Now that the basic model has been built, it is possible to compare it with the results of earlier works (Figure \ref{fig:crmnuss}). Although there are many processes at play in a full model, it may be possible to explain several of the differences between the two models. The \ion{C}{ii} ionisation cross-section used by \citeauthor{nussbaumer1975} is from the approximation of \citet{lotz1968}, which is noticeably below experiment for this particular ion. They also included fewer excitations to auto-ionising states above the threshold, leading to further differences with this work in the ionisation rates of \ion{C}{ii} and \ion{C}{iii}. \citeauthor{nussbaumer1975} could also be different with these ions and for \ion{C}{iv} because they use the general formula of \cite{burgess1969} for DR. Density effects on the population of \ion{C}{iv} can certainly be seen in the work of \citet{summers1974}, but it is being formed at an higher temperature than \citeauthor{nussbaumer1975} and this work. It suggests the model may have used either slower ionisation or faster recombination rates, or a combination of both. Comparison with the later ADAS modelling, which updated the work of \citeauthor{summers1974} and uses both the DR rates and modelling described in \citet{badnell2003}, shows similar results to the current work.

\begin{figure}
	\centering
	\includegraphics[width=9.4cm]{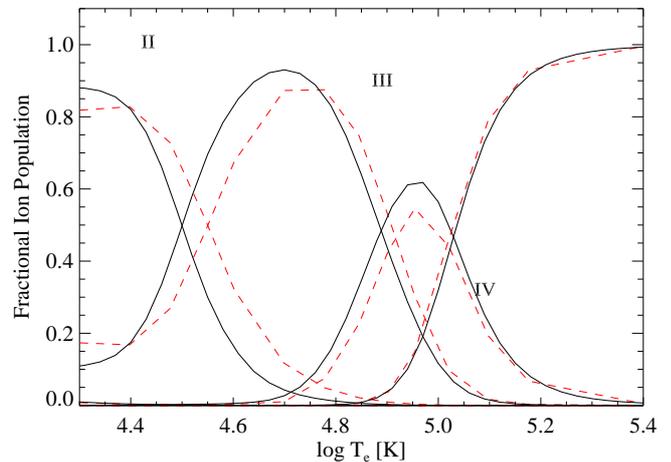}
	\caption{Comparison of the CR model and other work at $10^{10}$ cm$^{-3}$ density when photo-ionisation is included; black solid line - this work, red dashed - \citeauthor{nussbaumer1975}.}
	\label{fig:crmnusspi}
\end{figure}

Comparing the case with photo-ionisation included (Figure \ref{fig:crmnusspi}), differences in radiation field and PI cross-sections can reasonably be expected to produce variations between the two models; \citeauthor{nussbaumer1975} used radiances from a quiet Sun where some active regions were present. Allowing for this, it is possible to note the similarities in the overall charge-state distribution, especially considering the difference in \ion{C}{ii} collisional-ionisation rates between the two models, as seen above in the case without PI.

\section{Modelling the quiet-Sun transition region}
\label{sec:obs}

In order to test how much the present ion populations could affect plasma diagnostics, a means of comparing the results with observations is now discussed. One diagnostic technique involves predicting the line intensities using a given set of ion populations, and comparing them with actual line intensities.
	
Since the conditions to which this CR model most applies is for the solar TR, tests against the line emission from that region will be made. The technique used to calculate the theoretical line intensities is known as differential emission measure (DEM) modelling, (as described e.g. by \citealt{delzanna_mason:2018}). It uses the observed line intensities to estimate the amount of plasma present along the line of sight. The technique can be used in situations where the temperature distribution of the plasma varies smoothly along the field of view, which is a reasonable assumption for these solar conditions. It uses ion and level populations from the modelling, radiative decay rates and solar elemental abundances, along with the plasma density distribution with height, in order to calculate the expected intensity of each line. The DEM technique is also used for other purposes, such as determining elemental abundances, for instance.

The DEM modelling is dependent on ion populations, which is the very quantity being tested here. As a result, ion populations which have been previously published should be used to calculate the density distribution of the plasma with height based on observed line intensities. Once this has been established, the predicted intensities using the carbon ion populations from the different types of modelling can be calculated, and compared with observations.

\subsection{Method to compare the CR model with observations}

The modelling of the quiet-Sun TR observed on-disc is considered. The present analysis differs from the one in \cite{delzanna2019}, as EUV irradiances in the 60--1040~\AA\ range were used there. It also differs from the analysis presented in \cite{parenti_etal:2019}, as data from a narrow UV range obtained by the SoHO Solar Ultraviolet Measurement of Emitted Radiation (SUMER) instrument was used. The radiances used to calculate the photo-ionisation rates in Section \ref{sec:pi} are not used for the DEM because they do not extend to wavelengths which are important for carbon line emission in the TR.

As a baseline, the quiet-Sun UV and EUV radiances obtained with the 
Skylab SO55 Harvard instrument by \cite{vernazza_reeves:1978} are used here because of the more extended spectral range. The 35\% radiometric uncertainty and spectral resolution of about 1.6~\AA\ of the instrument are noted. Only a few, relatively unblended, strong lines are selected, formed at temperatures close to those of \ion{C}{ii} - \ion{C}{iv}. Much better spectra in the UV (in terms of spectral resolution and radiometric calibration) were obtained by SUMER. However, as pointed out by \cite{wilhelm1998a}, very good agreement between the SUMER and Skylab SO55 quiet-Sun radiances is present when lines relatively bright and isolated are considered. The Skylab radiances have therefore been supplemented with a few among those reported by \cite{wilhelm1998a} for the quiet Sun during the 1996 solar minimum. Often, good agreement is found with the quiet-Sun radiances obtained by OSO-4 \citep{dupree:1972}. As shown by \citet{andretta_delzanna:2014} and \citet{delzanna_andretta:2015}, solar-cycle variations of quiet-Sun radiances and irradiances for TR lines are smaller than the calibration uncertainties. For weaker or unresolved lines the quiet-Sun SUMER radiances of \citet{parenti_etal:2005a} have been considered, although it can be seen that some noticeable differences with both the Skylab SO55 and SUMER radiances of \cite{wilhelm1998a} are present.

Chianti v.8 atomic data \citep{delzanna_chianti_v8} was used and the \cite{asplund_etal:09} photospheric abundances. The DEM modelling was carried out with the improved version released in Chianti v.9 \citep{dere2019}. It uses MPFIT to find the best solution, assuming the DEM is a spline distribution.

The intensity of the vast majority of the strongest spectral lines is largely independent of density. However, some TR lines are sensitive to the varying density in the lower part of this region. It is customary to model TR lines with constant pressure, and the value selected for this work was 3$\times$10$^{14}$ cm$^{-3}$ K$^{-1}$, obtained from the model atmosphere given by \citet{avrett2008}. The effective temperature of a line is given by:

$$
T_{\rm eff} = \int G{\left({T}\right)}~
DEM{\left({T}\right)} ~T~dT~ /~
{\int G{\left({T}\right)}~DEM{\left({T}\right)}~dT}~,
$$

\noindent which is an average temperature more indicative of where a line is formed. As the DEM increases exponentially towards lower temperatures, $T_{\rm eff}$ is normally much lower than $T_{\rm max}$, the temperature where the contribution function $G(T)$ of a line has a maximum. In turn, $T_{\rm max}$ is normally higher than the temperature of maximum ion population in equilibrium.

\subsection{Results of the DEM modelling}

As a baseline, the zero-density Chianti charge-state distributions for all ions were used to determine the DEM from the observed lines. Carbon lines were excluded from this process so that the results for carbon from zero-density modelling and this work are treated equally; it also prevents the zero-density modelling of Li-like \ion{C}{iv} adversely affecting the DEM. Once the DEM was determined, the ratio of predicted to observed intensities for all of the lines were calculated. The ratios using the Chianti ion populations are shown in column $R_{\rm1}$ of Table \ref{tab:lines1}, and the ratios using the ion populations from this work are given in column $R_{\rm2}$.

\begin{table}[!htbp]
	\caption{Comparison of predicted and observed quiet-Sun UV and EUV radiances.} 
	\begin{center}
		\begin{tabular}{@{}lrrcccc@{}}
			\hline\hline \noalign{\smallskip}
			Ion & $\lambda_{\rm obs}$ & $I_{\rm obs}$ & $T_{\rm max}$ & $T_{\rm eff}$ & $R_1$ & $R_2$ \\
			\noalign{\smallskip}\hline\noalign{\smallskip}
			
			\ion{S}{ii} & 1253.80 & 15.9 &  4.50 &  4.41 &  1.05 &  \\
			\ion{S}{ii} & 1102.30 & 4.4 &  4.55 &  4.44 &  1.08 &  \\
			\ion{N}{ii} & 1085.70 & $36.7^a$ &  4.65 &  4.49 &  1.34 &  \\
			\ion{N}{ii} &  915.60 & 5.0 &  4.70 &  4.53 &  0.77 &  \\
			\ion{Si}{iii} & 1206.50 & 694.6 &  4.80 &  4.57 &  0.82 &  \\
			\ion{S}{iii} & 1200.96 & $8.0^b$ &  4.80 &  4.64 &  1.09 &  \\
			\ion{N}{iii} &  991.60 & 47.2 &  4.95 &  4.88 &  0.73 &  \\
			\ion{N}{iii} &  685.70 & 23.7 &  4.95 &  4.95 &  1.13 &  \\
			\ion{O}{iii} &  702.70 & 56.6 &  5.00 &  4.98 &  0.76 &  \\
			\ion{O}{iii} &  525.90 & 23.2 &  5.05 &  5.03 &  0.78 &  \\
			\ion{S}{iv} &  753.70 & 1.4 &  5.05 &  5.05 &  0.99 &  \\
			\ion{S}{iv} &  661.40 & 6.9 &  5.05 &  5.06 &  1.10 &  \\
			\ion{O}{iv} &  554.00 & 159.5 &  5.20 &  5.20 &  1.10 &  \\
			\ion{Ne}{iv} &  544.00 & 9.1 &  5.25 &  5.25 &  0.72 &  \\
			\ion{O}{v} &  630.00 & 335.0 &  5.40 &  5.38 &  1.00 &  \\
			\ion{Ne}{v} &  572.10 & 8.8 &  5.45 &  5.48 &  1.01 &  \\

			\noalign{\smallskip} \hline \noalign{\smallskip}
			
			\ion{C}{ii} & 1335.70 & 1205.0 &  4.60 &  4.45 &  0.92 & 0.71 \\
			\ion{C}{ii} & 1036.30 & $35.9^a$ &  4.65 &  4.49 &  1.92 & 1.29 \\
			\ion{C}{iii} &  977.00 & $702.0^a$ &  4.95 &  4.82 &  0.67 & 1.07 \\
			\ion{C}{iii} & 1176.37 & $36.2^a$ &  4.95 &  4.82 &  0.67 & 1.01 \\
			\ion{C}{iii} & 1175.74 & $104.0^a$ &  4.95 &  4.82 &  0.70 & 1.06 \\
			\ion{C}{iii} & 1174.88 & $37.4^a$ &  4.95 &  4.82 &  0.65 & 0.98 \\
			\ion{C}{iv} & 1548.20 & $361.0^a$ &  5.05 &  5.07 &  0.30 & 0.68 \\
			
			\noalign{\smallskip}\hline
		\end{tabular}
	\end{center}
	\normalsize
	\label{tab:lines1}
	\tablefoot{Ion - principal ion emitting at observed wavelength; $\lambda_{\rm obs}$ (\AA) - the observed wavelength; $I_{\rm obs}$ - the measured radiance (ergs cm$^{-2}$ s$^{-1}$ sr$^{-1}$) using \citeauthor{vernazza_reeves:1978}, except those marked by superscript: a) using \citeauthor{wilhelm1998a}, b) using \citeauthor{parenti_etal:2005a}; $T_{\rm max}$ and $T_{\rm eff}$ - the maximum and effective temperature using Chianti ionisation equilibria (log values, in K - see text); $R_1$ - the ratio between the predicted and observed intensities using populations from Chianti for all elements; $R_2$ - the same ratio using carbon populations from this work.}
\end{table}

Regarding the results of Chianti in column $R_{\rm 1}$, clearly there are some issues at lower temperatures, particularly with \ion{N}{ii} and \ion{N}{iii}, where the ratios of predicted to observed intensities for lines emitted within the same ion are not comparable with each other. Since both pairs of lines from these nitrogen ions are largely independent of density, it suggests the cause is more likely to be related to the formation temperatures of the lines. Based on the evidence of modelling carbon, it is reasonable to infer that density-dependent modelling would shift the formation temperatures of nitrogen, and provide improved theoretical values. It is also noted that the \ion{N}{ii} measurements do not come from the same observations, although this does not apply to the \ion{N}{iii} measurements, which show the same discrepancy. Although for \ion{O}{iii} lines the predicted intensities are lower than observations, both ratios are similar. For the lines which are formed at higher temperatures, the zero-density modelling clearly provides better results than at lower temperatures, which confirms the discussion in Section \ref{sec:intro}.

Looking at the carbon ions, \ion{C}{ii} is particularly interesting. Some opacity effects are present as both ratios of the strong multiplets around 1036 and 1335~\AA\ deviate from the theoretical ones in the optically-thin approximation. However, the deviations are not large and are similar. For example, the ratio of the 1335.7 (self-blend) and 1334.5~\AA\ radiances is about 1.4 as measured by SUMER \citep[cf.][]{judge_etal:2003} or 1.3 as measured by HRTS \citep[][]{sandlin_etal:86}, compared to a theoretical ratio of nearly two. The ratio of the 1037 and 1036.3~\AA\ doublet is about 1.3 as measured on-disc by SUMER \citep{wilhelm1998a}, compared to the optically-thin ratio of two. The \ion{C}{ii} lines are nearly Gaussian at Sun-centre, but become increasingly broad and affected by opacity towards the solar limb. 

The ratio of the line with lower oscillator-strengths, at $1036.3~\AA$, with the 1335.7~\AA\ self-blend is strongly temperature-sensitive. With an isothermal approximation, the observed radiances indicate a temperature of log $T$[K] = 4.23, that is, lower than the temperature of line formation in equilibrium with Chianti, log $T$[K] = 4.45, and with this work, log $T$[K] = 4.40. Indeed, there is a significant discrepancy in this ratio of 2.1 as predicted using the Chianti tables. Using the ion fractions from the level-resolved modelling reduces the ratio to 1.8.

Considering \ion{C}{iii}, as described for example in \cite{delzanna_mason:2018}, the ratio of the resonance line at 977~\AA\ with any line within the 1175~\AA\ multiplet is both density and temperature sensitive. With the results both from this work and Chianti, the ratios of the lines within the ion are close to one another, suggesting the pressure used in the modelling is a good reflection of the observed conditions. However, with the Chianti results, there is obviously a discrepancy with the predicted intensities in comparison to observations. Nevertheless, very good agreement with observation is obtained when using the ion populations modelled in this work.

The combined effect of lowering the formation temperature and the increased peak abundance of Li-like \ion{C}{iv}, owing to the density effects seen through the CR modelling, increases its predicted line intensity by more than a factor of two compared to the zero-density modelling, significantly reducing the discrepancy between predicted and observed intensities in the quiet Sun. 
	
It is noted that the DEM was established using zero-density ion populations. As discussed earlier, this does not fully reflect the conditions of higher density, lower temperature plasmas and could be affecting the determination of the DEM. Once ion populations for other elements have been modelled in the way described here, it may produce improvements in the DEM modelling and further agreement with observations.

\section{Conclusions}

This study has again highlighted, for carbon, in astrophysical plasmas of higher density, the importance of electron-impact ionisation from metastable levels and the suppression of dielectronic recombination in establishing the ion populations. The effect of ionisation from metastable levels reaches saturation once the level populations reach statistical equilibrium at high enough densities, whereas DR suppression makes an increasing contribution as density rises. Outer-shell excitation--auto-ionisation for \ion{C}{ii} and \ion{C}{iii}, which appears to have been neglected in many works, has been shown to contribute significantly to the shift with density in the formation temperatures of \ion{C}{iii} and \ion{C}{iv}.

Suppression of DR has been approximated as a way of determining its influence on the charge-state distribution. As stated in \citet{nikolic2013,nikolic2018}, the method is to be used as a guide only to determine whether full modelling is required. The work carried out here has shown that, in these plasma conditions, building the full CR model is not only warranted, but necessary in order to effectively model the ion populations. The focus of the work will now be to build a full model by the inclusion of high-lying levels. Initial indications show that photo-ionisation could also be important for the low charge-states of carbon, but the true extent of this will not be known until radiative transfer has been included. 

Modelling with the new atomic rates has improved predictions for the carbon charge-states observed in the quiet Sun, and in particular for Li-like \ion{C}{iv}. The results of the modelling are particularly relevant for understanding the formation of the carbon TR lines observed by missions like IRIS and the forthcoming Solar Orbiter SPICE spectrometer. It has been demonstrated that diagnostics derived from such missions which are dependent on ion populations require level-resolved charge-state distributions in order to correctly infer the plasma conditions.

\begin{acknowledgements}

Support by STFC (UK) via the consolidated grant of the DAMTP astrophysics group at the University of Cambridge is acknowledged, and the support of a University of Cambridge Isaac Newton Studentship.

The authors would like to thank the anonymous reviewer for highlighting areas where explanations could be clarified and the focus of the paper improved.

Most of the atomic rates used in the present study were produced by the UK APAP network, funded by STFC via several grants to the University of Strathclyde. Acknowledgent is made of the use of the OPEN-ADAS database, maintained by the University of Strathclyde.

Chianti is a collaborative project involving George Mason University, 
the University of Michigan, the NASA Goddard Space Flight Centre (USA) and the University of Cambridge (UK).

This document has been typeset using \LaTeX.

\end{acknowledgements}


\bibliographystyle{aa}

\bibliography{solartr_ionpop}

\end{document}